\def\bq{\begin{equation}}
\def\eq{\end{equation}}
\def\bqa{\begin{eqnarray}}
\def\eqa{\end{eqnarray}}
\def\bqb{\begin{eqnarray*}}
\def\eqb{\end{eqnarray*}}
\documentstyle[epsf,12pt]{article}
\def\pr#1#2#3{ Phys. Rev. ${\bf{#1}}$ (#2) #3}
\def\prl#1#2#3{ Phys. Rev. Lett. ${\bf{#1}}$ (#2) #3}
\def\pl#1#2#3{ Phys. Lett. ${\bf{#1}}$ (#2) #3 }

\def\np#1#2#3{ Nucl. Phys. ${\bf{#1}}$ (#2) #3}
\def\zp#1#2#3{ Z. Phys. ${\bf{#1}}$ (#2) #3}

\global\nulldelimiterspace = 0pt
 
\def\Bsl{\hbox{/\kern-.6700em$B$}} 
\def\Dsl{\hbox{/\kern-.6700em$D$}} 
\def\Wsl{\hbox{/\kern-.6700em$W$}} 

\def\roughly#1{\mathrel{\raise.3ex
    \hbox{$#1$\kern-.75em\lower1ex\hbox{$\sim$}}}}
\def\lsim{\roughly<}

\def\mh2{m^2_H}

\begin{document}
\pagenumbering{arabic}
\thispagestyle{empty}
\hspace {-0.8cm} hep-ph/9601306\\
\hspace {-0.8cm} PM/96-05\\
\hspace {-0.8cm} CPT-96/P.3304\\
\hspace {-0.8cm} January 1996\\
\vspace {0.8cm}\\
 
\begin{center}
{\Large\bf  Hadrophilic  $Z'$:} \\
\vspace {0.1cm}
{\Large\bf a bridge from LEP1, SLC and CDF to LEP2 anomalies} \\
 
 \vspace{1.8cm}
{\large  P. Chiappetta$^a$, J. Layssac$^b$, F.M. Renard$^b$ and C.
Verzegnassi$^c$}
\vspace {1cm}  \\
$^a$Centre
de Physique Th\'{e}orique,
UPR 7061,\\
CNRS Luminy, Case 907, F-13288 Marseille Cedex 9.\\
\vspace{0.2cm}
$^b$Physique
Math\'{e}matique et Th\'{e}orique,
CNRS-URA 768,\\
Universit\'{e} de Montpellier II,
 F-34095 Montpellier Cedex 5.\\
\vspace{0.2cm}
$^c$ Dipartimento di Fisica,
Universit\`{a} di Lecce \\
CP193 Via Arnesano, I-73100 Lecce, \\
and INFN, Sezione di Lecce, Italy.\\
 
\vspace{1.5cm}

 {\bf Abstract}
\end{center}
\noindent
In order to explain possible departures from the Standard Model 
predictions for $b\bar b$ and $c\bar c$
production at $Z$ peak, we propose the
existence of a $Z'$ vector boson with enhanced couplings to quarks. We
first show that this proposal is perfectly consistent with the full set
of LEP1/SLC results. In particular, $Z-Z'$ mixing effects naturally
explain the fact that $\Gamma_b$ and $\Gamma_c$ deviate from the SM in
opposite directions. We then show that there is a predicted 
range for enhanced $Z'q\bar q$ couplings which explains, for a precise
and interesting range of $Z'$ masses, the excess of dijet events seen
at CDF. A $Z'$ with such couplings and mass would produce clean 
 observable effects in $b\bar b$ and in total hadronic production at
LEP2.

\vspace{1cm}

\setcounter{page}{0}
\def\thefootnote{\arabic{footnote}}
\setcounter{footnote}{0}
\clearpage
 
\section{Introduction}

The precision measurements in the leptonic sector at LEP1/SLC agree with 
the Standard Model (SM) predictions  at the level of a few permille
\cite{LP95}, which leads to drastic constraints on any type of New
Physics (NP) manifestation. As of today, the situation in
the quark sector is slightly different. Through measurements of
the $Z\to b\bar b$ and $Z\to c\bar c$ widths and asymmetries, 
LEP and SLC have given indications for possible departures
from the SM predictions for $b$ and $c$ couplings at the level of a few
percent. In the $b\bar b$ case  such anomalies could be interpreted
as a signal for NP in the heavy quark sector, driven for example
by the large value of the top mass, whose effects already appear at
standard level \cite{mt2SM}. Several models of this type have been
proposed (anomalous top quark properties \cite{Yuan}, \cite{GRV}, 
ETC models \cite{ETC}, anomalous gauge
boson couplings \cite{RVAGC}, supersymmetric contributions, 
new Higgses, gauginos,..\cite{SUSY}, \cite{Comelli}).
A common feature of all these explanations is that they fail to explain
the possible existence  of $c\bar c$ anomalies,
which cannot be enhanced by the large top mass. So it seems more
difficult to describe the presence of anomalies in both $b\bar b$ and
$c\bar c$ channels, without drastically modifying 
the fermionic sector, for
example through the mixing of quark multiplets with higher fermion
representations as proposed in\cite{Ma}.\par
In this paper we would like
to propose a simple explanation based on the existence of a 
\underline{hadrophilic Z'}
 vector boson, i.e. one which would couple universally to quarks more
strongly than to leptons. We shall not propose
here a specific model, although the concept of $Z'$ differently coupled
to quarks and to leptons has already been considered in the past
\cite{Georgi}. We shall be limited to extracting from  LEP1/SLC
experiments several suggestions about the required $Z'$ properties. To
achieve this, we shall first rely on a model independent framework for
the analysis of $Z-Z'$ mixing effects. This is available from a previous
work \cite{LRV} in which the $Z'$ couplings to each fermion-antifermion
pair were left free. Working in this spirit, we will then   
derive in section 2 experimental informations on the $Z-Z'$ mixing angle
$\theta_M$ and on the $Z'f\bar f$ couplings showing that, indeed, the
anomalies in $b\bar b$ and
$c\bar c$ production can be described by such an hadrophilic $Z'$. In
particular, from the absence of anomaly in the total hadronic width
$\Gamma_{had}$ at $Z$ peak \underline{we shall explain in a natural 
way the fact that the SM
departures in $\Gamma_b$ and in $\Gamma_c$ have}\\ 
\underline{opposite signs}.\par
The next relevant question to be answered is that of whether the values 
of the $Z'$ couplings that we determined in  this way do not contradict 
any already available experimental constraint.
In particular, we shall focus in section 3 on the significant excess of 
dijet events for large masses (above $500$ GeV) at CDF \cite{CDF}. We
shall show  that this phenomenon could be  naturally explained in terms
of an   hadrophilic
$Z'$, whose mass lies in the range between $800$ GeV and $1$ TeV and 
whose couplings are restricted by the request that the $Z'$ behaves
like a not too wide resonance, identifiable in different
processes.\par
 Our second step will then consist of examining in section 4 the
consequences of this solution for other processes, in particular possible
$Z'$ effects in $e^+e^-\to f\bar f$ at LEP2.\par Here the natural
final channels to be considered in our case are the hadronic ones, where
the $Z'$ effect would depend on the product of $Z'$ couplings to
leptons times $Z'$ couplings to quarks. In this paper, we shall
consider the pessimistic case where the leptonic $Z'$ couplings are
not sufficiently strong to give rise to visible effects in the leptonic
channel. Starting from this conservative assumption, we shall show
that it would be still possible  to observe effects in hadronic
channels. We will proceed in two steps. First, in a model-independent
way, we shall establish the domain of $Z'b\bar b$ couplings that would
lead to visible deviations in the $b\bar b$ cross section $\sigma_b$ and
in the forward backward asymmetry $A^b_{FB}$. We shall show that this
domain largely overlaps with the ones suggested by our analysis of
LEP1/SLC and CDF results. We shall then examine the total hadronic
cross section $\sigma_{had}$ at LEP2 and we shall find again
that the domain of $Z'b\bar b$ and $Z'c\bar c$ couplings leading to
visible effects contains the values selected by LEP1/SLC and CDF.\par 
We can
therefore conclude that, if a hadrophilic $Z'$ is at  the origin of the
present observed anomalies, a quantitative study of these three hadronic
observables at LEP2 would allow to confirm this relatively simple
explanation. In this case, it would become relevant and meaningful to
construct a full and satisfactory theoretical model.\par

\newpage

\section{Analysis of LEP1/SLC results in terms of $Z-Z'$ mixing.}

We consider $Z-Z'$ mixing effects at the $Z$ peak in a model independent
way following the procedure given in ref.\cite{LRV}. As well-known, the
two relevant effects consist in a modification of the $Z$ couplings to
fermions, proportional to a mixing angle $\equiv \theta_M$, and in a $Z$
mass shift which induces a contribution to the $\delta_{\rho}$
parameter:

\bq  \delta^{Z'}_{\rho} \simeq \theta^2_M {M^2_{Z'}\over M^2_Z} \eq

The quantity $\delta^{Z'}_{\rho}$ is a \underline{positive quantity} 
that can be extracted from the ratio 
$c_w^2 \equiv {M^2_W\over M^2_Z}$ and its comparison to the
quantities measured at the $Z$ peak 
 and defined in the conventional way \cite{epsilon}.

From the latest available data \cite{LP95} and
under the assumption that no other  significant contributions to 
$\delta_{\rho}$ (e.g. from one extra $W'$) exists, we obtain at two
standard deviations:

\bq  0 \leq \delta^{Z'}_{\rho} \leq +0.005  \label{rho} \eq
\noindent
In this way we derive an upper value for the mixing angle:

\bq  |\theta_M| < \sqrt{0.005} {M_Z\over M_{Z'}} \label{tmmax}  \eq
\noindent
Note that for our nextcoming qualitative analysis,  values of 
$\delta^{Z'}_{\rho}$ not unreasonably larger than the limit of
eq.~(\ref{rho}) would not modify our conclusions. We shall come back
on this point later. We then normalize the $Z'f\bar f$ couplings:

\bq  -i{e(0)\over 2s_1c_1}\gamma^{\mu}[g'_{Vf}-g'_{Af}\gamma^5]  \eq 

\noindent
in the same way as the $Zf\bar f$ ones:

\bq -i{e(0)\over 2s_1c_1}\gamma^{\mu}[g_{Vf}-g_{Af}\gamma^5] \eq 
\noindent
with $g_{Vl}=-{v_1\over2}$; $g_{Al}=-{1\over2}$;
$g_{Vf}=I^3_f-2s^2_1Q_f$; $g_{Af}=I^3_f$; 
$v_1=1-4s^2_1$; 
$s^2_1\equiv1-c^2_1\simeq 0.2121$ 
from $s^2_1c^2_1=\pi\alpha(0)/\sqrt2 G_{\mu}M^2_Z$.\par
This allows us to define the ratios:

\bq \xi_{Vf} \equiv {g'_{Vf}\over g_{Vf}}   \ \ \ \ \ \ \ \ \ \ 
\xi_{Af} \equiv {g'_{Af}\over g_{Af}}   \eq
\noindent
which will significantly measure the magnitude of the $Z'f\bar f$ 
couplings.
Keeping in mind the fact that $g_{Vl}$ is depressed by
$v_1\simeq0.1516$,
we will consider as "natural" (i.e. non enhanced) magnitudes
$\xi_{Al}\simeq1$, $\xi_{Vf}\simeq1$, $\xi_{Af}\simeq1$ for $f\neq l$, 
but $\xi_{Vl}\simeq6$.\par
The total fermionic $Z'$ width is given by

\bq  \Gamma^{ferm}_{Z'} ={\alpha M_{Z'}\over 12s^4_1c^4_1}\sum_f
N_f(1-{4m^2_f\over M^2_{Z'}})^{1/2}[\xi^2_{Vf}g^2_{Vf}(1+{2m^2_f
\over  M^2_{Z'}})+\xi^2_{Af}g^2_{Af}(1-{4m^2_f\over  M^2_{Z'}})] 
\label{gamzp}  \eq

\noindent
$N_f$ being the lepton ($=1$) or quark ($=3$) colour factor.

The $Z-Z'$ mixing effects on Z peak observables ($Z$ partial widths
and asymmetries), due to $\delta^{Z'}_{\rho}$ and to the modifications
of the $Z$ couplings (of the form $\theta_M g'_{V,A}$) are analyzed
in Appendix A. Using the most recent LEP and SLC data \cite{LP95} we
obtain informations on $Z'$ couplings. They are summarized below in the
form of allowed bands, at two standard deviations, assuming that
$|\theta_M|$ saturates the bound,eq.~(\ref{tmmax}), 
(so in a sense these are minimal bands) with the two possible signs
$\eta_M=\pm1$.\par
\underline{$Z'l\bar l$ couplings}
 
\bq \eta_M \xi_{Vl} \simeq (-2.25 \pm 6.25)({M_{Z'}\over 1 TeV}) \ \ \
(LEP) \ \ \ \ \ \ \ \ \
\eta_M \xi_{Vl} \simeq (+1.75 \pm 6.25)({M_{Z'}\over 1 TeV}) \ \ \
(SLC) \eq

\bq  \eta_M \xi_{Al} \simeq (-0.2 \pm 0.5)({M_{Z'}\over 1 TeV})  \eq

\underline{$Z'b\bar b$ couplings}

\bq \eta_M \xi_{Vb} \simeq (-3.45 \pm 20.72)({M_{Z'}\over 1 TeV}) \ \ \
(LEP) \ \ \ \ \
\eta_M \xi_{Vb} \simeq (-24.24 \pm 25.98)({M_{Z'}\over 1 TeV}) \ \ \
(SLC) \label{lep} \eq

\bq \eta_M \xi_{Ab} \simeq (+4.58 \pm 9.84)({M_{Z'}\over 1 TeV}) \ \ \
(LEP) \ \ \ \ \ \
\eta_M \xi_{Ab} \simeq (+14.54 \pm 12.47)({M_{Z'}\over 1 TeV}) \ \ \
(SLC) \label{slc} \eq

\underline{$Z'c\bar c$ couplings}

\bq \eta_M \xi_{Vc} \simeq (-6.94 \pm 26.60)({M_{Z'}\over 1 TeV}) \ \ \
(LEP) \ \ \ \ \ \
\eta_M \xi_{Vc} \simeq (-20.38 \pm 40.62)({M_{Z'}\over 1 TeV}) \ \ \
(SLC) \eq

\bq \eta_M \xi_{Ac} \simeq (-7.88 \pm 8.46)({M_{Z'}\over 1 TeV}) \ \ \
(LEP) \ \ \ \ \ \
\eta_M \xi_{Ac} \simeq (-6.01 \pm 9.70)({M_{Z'}\over 1 TeV}) \ \ \
(SLC) \eq

Because of the various uncertainties, both theoretical (the assumption
about $|\theta_M|$) and experimental (disagreements for
various measurements and large errors in the quark cases) we take these
results just as indicative and we call the resulting values
\underline{suggested $Z'$ couplings}. Several important remarks are
nevertheless in order.\par 
First, as expected, lepton couplings are
strongly constrained: $\xi_{Vl}$ and $\xi_{Al}$ lie within the "natural"
range mentioned above.\par Secondly,on the contrary, there is room for
very large values for quark couplings. In one case, from SLC data, a
definite non zero value for $\xi_{Ab}$ is suggested. Obviously the
extreme quoted values are to be taken as purely indicative. A priori we
would not trust values larger for example than the QCD strength (
$\alpha_s\simeq0.12$), which implies
$|\xi_{Af}|<7$ and $|\xi_{Vf}|<7/v_f$, i.e. $|\xi_{Vb}|<10$ and
$|\xi_{Vc}|<16$. We will \underline{conventionally} define as
 "reasonable" the values of the couplings lying within this range.
Further restrictions can a priori be set by considering their
effects on the total fermionic $Z'$ width eq.~(\ref{gamzp}). 
This will be
discussed in the next section.\par

There is one more important information to be extracted from $Z-Z'$
mixing effects at $Z$ peak. From the very precise measurement of
$\Gamma_{had}$ leading to:

\bq  {\delta \Gamma_{had}\over \Gamma_{had}} =+0.003\pm 0.0017  
\label{gamlep} \eq
and eq.(A.7) one obtains

\bq  \eta_M[4v_c\xi_{Vc} + 12\xi_{Ac} +12v_b\xi_{Vb} +18\xi_{Ab}] =
(10.6\pm15.4)({M_{Z'}\over 1 TeV}) \label{correl} \eq

\noindent
where $v_f=1-4|Q_f|s^2_1$. In practice, up to a small uncertainty, this
relation reduces the 4-parameter quark case to a 3-parameter one.
This result, valid for the most general type of $Z'$, will introduce
a quite useful simplification in our nextcoming calculations.\par
 From eq.~(\ref{gamlep})  we can derive a strong correlation between 
$\delta \Gamma_b$ and $\delta \Gamma_c$, that is peculiar of our $Z'$ 
hypothesis. Our universality assumptions 
$\delta^{Z'} \Gamma_u = \delta^{Z'} \Gamma_c$ and 
$\delta^{Z'} \Gamma_d = \delta^{Z'} \Gamma_s = \delta^{Z'} \Gamma_b$
allow us to rewrite eq.~(\ref{gamlep}) as:
\bq  {\delta \Gamma_{had}\over \Gamma_{had}} = 
2 ({{\delta \Gamma_c}\over \Gamma_c}) ({\Gamma_c\over \Gamma_{had}})
 + 3 ({{\delta \Gamma_b}\over \Gamma_b}) ({\Gamma_b\over \Gamma_{had}}) 
\eq
leading to the conclusion:
\bq {\delta \Gamma_b\over \Gamma_b} = - ({{2}\over{3}})
({{R_c}\over {R_b}}) ({\delta \Gamma_c\over \Gamma_c}) + 
({{1}\over{3
R_b}})
 ({\delta \Gamma_{had}\over \Gamma_{had}})  \eq
Numerically the second term of the right hand part is negligible in 
first approximation, which finally gives:

\bq {\delta \Gamma_b\over \Gamma_b} \simeq -0.5  
{\delta \Gamma_c\over \Gamma_c}
\eq

Thus, in a natural way, the relative shifts in $\Gamma_b$ and in 
$\Gamma_c$ are predicted to be of opposite sign, with a ratio
consistent with the experimental data and errors, which is a peculiar
feature of the model, valid for all the values of its quark couplings
that obey the universality request.

  Finally, note that the values of
these suggested $Z'$ couplings  grow linearly with the mass
$M_{Z'}$. This is a natural consequence of assuming a given 
$Z-Z'$ mixing effect on the $Z$ peak
observables. When $M_{Z'}$ grows, $\theta_M$ decreases. Consequently 
for a given $Z-Z'$ mixing effect the required $Z'$ couplings
increase.
\par  Our model independent analysis of the LEP1/SLC
constraints on the $Z'$  parameters is thus finished. In the next
section, we shall investigate whether the large "suggested" $Z'q\bar q$
couplings are not ruled out by the data available from the hadronic
colliders.

\newpage

\section{Analysis of CDF dijet events in terms of a $Z'$ resonance.}
The CDF collaboration has reported the observation of an excess of
events  with two-jet mass above $500$ GeV, compared to the QCD
prediction. The jets  have been required to satisfy $|\eta| < 2$
($\eta$ being the pseudorapidity) and the events are required to have 
$|\cos{\theta}^{\star}| < \frac{2}{3}$, ${\theta}^{\star}$ being
the parton scattering angle in the partonic center of mass frame. This
kinematical  restriction favors the appearance of NP since the QCD
cross section is peaked around $|\cos{\theta}^{\star}|\simeq 1$. The
two jet production in hadronic collisions has been computed at next to
leading order in QCD \cite{EKS}. The aim of this section is that of
investigating whether the observed dijet excess may, or may not, be
explained in terms of a hadrophilic $Z'$, that a priori represents in
our opinion a reasonably natural possibility. In order to pursue this
program we have to calculate the effect of the addition to the
dominant QCD component of the weak contribution.
In the SM this comes from W,Z and photon exchanges. In our analysis we
will add the extra contribution due to the $Z'$, with couplings taken
within the range suggested by the LEP/SLC analysis. The practical
calculation is rather lengthy and will be summarized in Appendix B.
\par The weak contribution being evaluated at leading order we
shall perform the calculation of the strong part at the same level.
It has been shown in \cite{EKS} that the difference between the
order $\alpha_s^3$ calculation and the Born calculation is small
provided that we fix the arbitrary factorization M and renormalization
$\mu$ scales to:
\bq  M= \mu = \frac{0.5 M_{JJ}}{2 \cosh(0.7\eta_{\star})} \label{scale}
\eq 
where $M_{JJ}$ is the dijet mass and 
$\eta_{\star}=\frac{|\eta_1 - \eta_2|}{2}$, 
$\eta_i$ being the pseudorapidity of jet i. In the following we will
use the prescription given in eq.~(\ref{scale}).
 The deviation from the QCD prediction appears as a resonance bump
in the $700-1000$ GeV $M_{JJ}$ mass range, suggesting therefore an
indicative $Z'$ mass range around $700-1000$ GeV. Since the bump is
wide, the hadrophilic $Z'$ cannot be narrow. \par  
The results of our investigation are shown in figures
1 and 2. As one can see, the observed dijet excess can be
satisfactorily explained for $M_{Z'}$ around  $800-900$ GeV and for
reasonable $Z'q\bar q$ values i.e.$|\xi_{Af}|$ and  $|\xi_{Vf}| \simeq 3$. 
We have checked that these values satisfy the correlation
constraint due to $\Gamma_{had}$,
eq.~(\ref{correl}) and lead to an acceptable enhancement of the $Z'$
width eq.~(\ref{gamzp}). 
Note that $|\xi_{Af}|$ and $|\xi_{Vf}|$ cannot be
simultaneously too small (i.e. all $\simeq 1-2$), otherwise the width
would be too narrow. To fix a scale in our analysis we allow the $Z'$
width to lie in the range $\Gamma_{Z'} \simeq 150-200$ GeV. Larger
values of the $Z'q\bar q$ couplings would lead to an unreasonably wide
resonance and the observed peak would be much less pronounced.\par The
excess of dijet events could also be explained by an hadrophilic $Z'$
of mass $M_{Z'}=700$ GeV or even $1$ TeV provided that its quark
couplings are all suitably larger, i.e. for $|\xi_{Af}|$ and
$|\xi_{Vf}|$ values between $3$ and $5$. For what concerns possible
effects at LEP2 these situations would lead to more dramatic
consequences. For this reason, we shall rather concentrate our analysis
on the configuration of figures 1 and 2, which corresponds from this
point of view to a more conservative attitude.\par A few technical
comments about our calculation are now appropriate. We have used the
KMRS set B of parton distributions \cite{KMRS}. The uncertainty due to
our imperfect knowledge of the structure functions is small since we
calculate a ratio. The dominant weak contribution is  due to the $Z'$
pole. We are therefore not sensitive to the sign of $Z'q\bar q$ couplings
and the SM weak vector bosons contributions are quite negligible in the
high dijet mass range we are interested into.\par
 This concludes our confrontation of hadrophilic $Z'$
hypothesis to existing data. We shall now investigate the future
prospects from LEP2.

\newpage
\section{ Z' Effects in hadronic production at LEP2}
In this section, we shall examine possible visible consequences of our
assumption that a hadrophilic $Z'$ exists, with "suggested" couplings
and mass derived by an overall analysis of LEP/SLC and CDF data. As
rather natural experimental quantities to be considered for this
purpose, we shall concentrate our attention on the three hadronic
observables that will be measured in a very near future at LEP2, i.e. 
the $b\bar b$ cross section $\sigma_{b}(q^2)$, the $b\bar b$ forward
backward asymmetry  $A_{FB,b}(q^2)$ and the total hadronic production
cross section $\sigma_{h}(q^2)$, where $\sqrt {q^2}$ is the total center
of mass energy that will vary in the range (chosen for theoretical and
experimental reasons \cite{LEP2}) $ 140 GeV \leq \sqrt {q^2} \leq 190
GeV$. The calculated shifts on these three quantities due to a $Z'$
will depend on products of $Z'$ quark couplings with $Z'$ lepton
couplings. For the latter ones, we have seen from our previous
investigation that no special "suggestion" exists that motivates some
anomalously large values. In fact, a more detailed investigation of
the constraints on the $Z'$ lepton couplings derived from LEP/SLC
would lead to the conclusion that $Z'$ signals in the leptonic channel
at LEP2 are not forbidden, but are also not specially encouraged. In
particular, in the \underline{extreme} configuration of a saturation
of the bound on $ |\theta_M|$, the lepton couplings would lie in a
domain which corresponds roughly to the domain of non observability
for the various leptonic observables at LEP2, which has been derived
very recently in another detailed paper \cite{LEP2ZP}. Following a
conservative attitude, we shall assume therefore that the leptonic $Z'$
couplings lie in the previous domain of non observability at LEP2.
With this input, we shall look for possible effects in the LEP2
hadronic channels, motivated by the suggested anomalously large $Z'$
quark couplings. Of course, should an effect be produced in the
leptonic channel, the corresponding situation in the hadronic one
would become more favourable than in the configuration that we shall
consider from now on.\par
The treatment of the $Z'$ shifts on various observables can be
performed in various ways. We shall follow in this paper a theoretical
approach that has been proposed very recently \cite{univ}, in which
this effect can be formally considered as a one loop $Z'$ correction
of "box" type to the SM quantities containing conventional $\gamma$
and Z exchanges. These corrections enter in a not universal way in
certain gauge-invariant combinations of self-energies, vertices and
boxes that have been called $\tilde{\Delta}\alpha(q^2)$, $R((q^2)$, 
$V_{\gamma Z}(q^2)$ and $V_{Z \gamma}(q^2)$, whose contributions to
the various observables have been completely derived and thoroughly
discussed in the section 2 of ref.\cite{univ}. We shall not repeat
here the derivation of these contributions, and defer the interested
reader to the aforementioned reference. For our purposes, it will be
sufficient to remind that the relevant one-loop corrected
expressions of an observable $O_{lf}$ of the process 
$e^+e^-\to f \bar f$ (where f is a certain quark) will be of the
type: 
 \bq O_{lf}(q^2) = O_{lf}^{(Born)} \lbrack 1 + a_{lf} 
\tilde{\Delta}^{(lf)}_{\alpha}(q^2) +  b_{lf} R^{(lf)}(q^2) + c_{lf} 
V^{(lf)}_{\gamma Z}(q^2) + d_{lf} V^{(lf)}_{Z \gamma}(q^2) \rbrack
 \eq

\noindent
where $(a,b,c,d)_{lf}$ are certain numerical constants given in
ref.\cite{univ} for the various relevant cases and $ O_{lf}^{(Born)}$
is a certain suitably defined "effective" Born approximation. For the
case $f=b$, the $Z'$ contributions to the four one loop corrections
turn out to be:
 
\bq \tilde{\Delta}^{(lb)}_{\alpha}(q^2) = -z_{2l}z_{2b} \ \ \ \ \ \ \ \ \ 
R^{(lb)}(q^2) = z_{1l}z_{1b}\chi^2 \label{sub1} \eq

\bq V^{(lb)}_{\gamma Z}(q^2) = z_{1l}z_{2b}\chi^2 \ \ \ \ \ \ \ \ \ 
V^{(lb)}_{Z\gamma}(q^2) = z_{2l}z_{1b}\chi^2 \label{sub2} \eq
\noindent

where we use the reduced couplings:
\bq  z_{1b} =  \xi_{Ab} \sqrt{{q^2\over M^2_{Z'}-q^2}} \eq

\bq  z_{2b} = ({3v_b\over
4s_1c_1})(\xi_{Vb}-\xi_{Ab})\sqrt{{q^2\over M^2_{Z'}-q^2}}  \eq
\noindent
and ${\chi}^2= \frac{(q^2- M^2_{Z})}{q^2}$.\par

 From these expressions  we have computed the relative shifts
${\delta\sigma_{b}(q^2)\over\sigma_{b}}$ and
${\delta A_{FB,b}(q^2)\over A_{FB,b}}$ due to a $Z'$, assuming as
previously discussed that the lepton couplings lie in the domain of
non observability at LEP2. As it has been shown in \cite{LEP2ZP}, this
corresponds to the following limitations on the leptonic ratios:

\bq   |\xi_{Vl}| \lsim ({0.22\over
v_1})\sqrt{{M^2_{Z'}-q^2\over q^2}} \label{vl}  \eq

\bq   |\xi_{Al}| \lsim (0.18)\sqrt{{M^2_{Z'}-q^2\over q^2}} \label{al}
\eq

The calculation of the shifts has been performed without taking into 
account the potentially dangerous effects of QED radiation. From our
previous experience \cite{LEP2ZP} we know that, provided that
suitable experimental cuts are imposed, the realistic results will
not deviate appreciably from those calculated without QED
convolution. This is particularly true if one is interested in large
effects, as in our case. We defer the reader to ref \cite{LEP2ZP}
for a complete disussion of this point. \par
 From now on, we shall
concentrate on the configuration  $q^2=(175 GeV)^2$ since, for the
purposes of $Z'$ searches, it has been shown in \cite{LEP2ZP} that
within the three planned realistic LEP2 phases this is the most
convenient one. In this case, we can rewrite for sufficiently large
$M_{Z'}$ (which we are assuming) eq.~(\ref{vl}) and  eq.~(\ref{al}) as:

\bq |\xi_{Vl}|\lsim 8.02({M_{Z'}\over 1 TeV}) \ \ \ \ 
|\xi_{Al}|\lsim 1.01({M_{Z'}\over 1 TeV})  \label{unseen} \eq
\noindent

\par
In figures 3 and 4 we present our results for the $Z' b \bar b$
couplings rescaled by the factor ${M_{Z'}\over 1 TeV}$. The
observability regions of figure 3 correspond to a relative $Z'$ effect 
in ${\delta\sigma_{b}\over\sigma_{b}}$ of at least five percent
(dark area) and ten percent (grey area). In figure 4, numerical
effects of five and ten percent on the relative forward-backward
asymmetry  ${\delta A_{FB,b}\over A_{FB,b}}$ are depicted.
Following the analysis presented in table 2 of ref.\cite{LEP2ZP},
these $Z'$ effects would be visible in the chosen LEP2 configuration.
Note that we have restricted the variation domain of variables in the
figures to values that we called "reasonable" in section 2, i.e. that 
contain in fact the strip $|\xi_{Ab}| = |\xi_{Vb}| \simeq 3$ suggested
by our previous CDF analysis. Note that we did not fix the $M_{Z'}$
value. To be consistent with our prefered CDF choice 
$M_{Z'} \simeq 800-900$ GeV, we should in fact rescale the values of
the couplings shown in the figures 3 and 4 by a (scarcely relevant) 
$10-20 \%$ factor. \par
As one can see from an inspection of the two figures, values of the
couplings lying in the neighbourhood of the "suggested" representative
set of couplings $|\xi_{Ab}| = |\xi_{Vb}| \simeq 3$ would produce
in both cases a large effect. In other words, a hadrophilic $Z'$ with
such couplings and mass should not escape indirect experimental
detection in the final $b \bar b$ channel at LEP2.\par

 We discuss now the possible $Z'$ effects on the total
hadronic cross section $\sigma_{had}$ (hereafter denoted $\sigma_5$)
at LEP2. 

For up quarks we use the reduced couplings: 

\bq  z_{1c} =\xi_{Ac}  \sqrt{{q^2\over M^2_{Z'}-q^2}}  \eq

\bq  z_{2c} =({3v_c\over
8s_1c_1})(\xi_{Vc}-\xi_{Ac}) \sqrt{{q^2\over M^2_{Z'}-q^2}}  \eq
\noindent
and the quantities corresponding to eq.~(\ref{sub1}) and
 eq.~(\ref{sub2}) with the replacement  of $b$ by $c$.
The expression of $\sigma_5(q^2)$ is taken from ref.\cite{univ}
and we considered the relative shift
${\delta\sigma_5\over \sigma_5}$ expressed in terms of the eight
quantities corresponding to eq.~(\ref{sub1}) and
 eq.~(\ref{sub2}) for up quarks ($c$) and down quarks ($b$). 
A priori they
depend on four $Z'$ couplings $\xi_{Vb}$, $\xi_{Ab}$, $\xi_{Vc}$, 
$\xi_{Ac}$. We imposed the strong correlation  eq.~(\ref{correl})
implied by the absence of effect in $\Gamma_{had}$, which practically
reduces the freedom to a small domain around a three independent quark
parameters case. As above we kept the
leptonic $Z'$ couplings inside the non observability domain at
LEP2, eq.~(\ref{unseen}).\par
With these inputs we looked for visible effects in $\sigma_5(q^2)$. 
The results are shown in figure 5, 
demanding ${\delta\sigma_5(q^2)\over \sigma_5}$ larger than 5\%. 
Following the experimental analysis of  ref.\cite{LEP2ZP}, this
relative shift would
represent a spectacular effect. 
One sees from this figure that indeed values of  couplings 
$|\xi_{Ab}| = |\xi_{Vb}| = |\xi_{Ac}| = |\xi_{Vc}|\simeq 3$, lying
around the suggested CDF ones, would be able to generate a clean and
impressive effect both in the  $b \bar b$ and in the total
hadronic observables. This would represent, in our opinion, a
spectacular confirmation of the $Z'$ origin of the apparent LEP/SLC
and CDF anomalies.

\section{Conclusions}

In order to explain  possible  $b \bar b$ and  $c \bar c$
anomalies observed in LEP1 and SLC experiments at Z peak, we used a
model independent description of $Z-Z'$ mixing effects starting with
arbitrary mixing angle and $Z'f\bar f$
couplings. With this description, using the full set of 
LEP1/SLC data at Z peak,
we have derived "suggested" $Z'$ couplings to leptons and quarks.
The presence of anomalous effects in hadronic
channels at $Z$ peak as opposed to very stringent 
constraints in leptonic channels would be explained by a 
$Z'$ more strongly coupled to quarks than to leptons,
a \underline{hadrophilic $Z'$}. We notice, as
a support to our assumption, that the absence of effect in $\Gamma_{had}$
leads naturally to the prediction of effects with opposite signs
in $\Gamma_b$ and in $\Gamma_c$, in agreement with experimental
data.\par
We considered the consequences of this hypothesis for other
processes. We have first investigated the observed excess of high mass
dijet events at CDF. This excess can be naturally explained by the
hadrophilic $Z'$ provided that its couplings to quarks are reasonable,
its mass range lies around $800-900$ GeV and its width is
relatively large ($\Gamma_{Z'} \simeq 200$ GeV).\par
  We have also examined the observability of
hadrophilic $Z'$ effects at LEP2. We have checked that for leptonic
channels, the "suggested" strongly constrained leptonic 
couplings do not particularly motivate $Z'$ effects at LEP2.\par
On the contrary  the suggested $Z'b\bar b$
couplings would produce large effects in $e^+e^-\to b\bar b$ 
(cross section and forward-backward asymmetry) at LEP2.
Within the assumption that the $Z'$ leptonic couplings are such that
no effect is seen in leptonic observables,
we have established
model independent observability domains in the space of vector and axial
$Z'b\bar b$ couplings. These domains correspond to
visible effects if the $Z'b\bar b$ couplings have a reasonably enhanced
magnitude.
There is a large overlap with the domains 
suggested by LEP1/SLC and CDF. So the existence
of a 
hadrophilic $Z'$ producing LEP1/SLC and CDF anomalies
could be confirmed by such measurements at LEP2.\par
We have
then analysed what information the total hadronic cross section could
bring on $Z'c\bar c$ couplings. The interesting feature is the strong
correlation imposed by the absence of effect in $\Gamma_{had}$ at $Z$
peak. With this constraint included in the analysis of $\sigma_{had}$
at LEP2, we have determined the observability domains in the space 
of vector and axial $Z'c\bar c$ couplings. We have established them in
correlation with
various ranges of "reasonable" $Z'b\bar b$ couplings. It appears that
visible effects would also be present in $\sigma_{had}$ for
similar "reasonable"
values of $Z'c\bar c$ couplings. Should this happen, a deeper
theoretical analysis on the origin of such an hadrophilic $Z'$
would become mandatory. \par

\vspace{0.5cm}

{\bf \underline{Acknowledgements}}\par
This work has been partially supported by the EC contract
CHRX-CT94-0579.
We thank C. Bourrely, J. Ph. Guillet and J.
Soffer for discussions. The analysis of the CDF anomaly in terms of a
$Z'$ was also independently performed in another publication
(G. Altarelli, N. di Bartolomeo, F. Feruglio, R. Gatto, M. Mangano,
CERN-TH/96-20, hep-ph9601324). One of us (C.V.) is indepted to G.
Altarelli for having suggested in the early stage of this work to
consider the role of the CDF dijet excess as a test of the hadrophilic
$Z'$ assumption.

 \newpage

\renewcommand{\theequation}{A.\arabic{equation}} \setcounter{equation}{0} %
\setcounter{section}{0}

{\large {\bf Appendix A : $Z-Z'$ mixing 
effects on $Z$ peak observables}}\\

\vspace{0.3cm}
 From the analysis of ref.\cite{LRV} we can derive the shifts to the SM
predictions for the various $Z$ peak observables, partial Z decay
widths ($\Gamma_f \equiv \Gamma(Z\to f \bar f)$ and asymmetry factors
$A_f$. Forgetting
systematically terms that are numerically negligible we get :

\bq  {\delta \Gamma_l\over \Gamma_l} = \delta^{Z'}_{\rho}
+ 2\theta_M \xi_{Al}  \eq

\bq  {\delta A_l} = 3\delta^{Z'}_{\rho}
+ 2\theta_M v_1 \xi_{Vl}  \eq

\bq  {\delta \Gamma_u\over \Gamma_u} = {8\over5}\delta^{Z'}_{\rho}
+ {3\over5}\theta_M [v_u\xi_{Vu} + 3\xi_{Au}]  \eq

\bq  {\delta \Gamma_d\over \Gamma_d} = {19\over13}\delta^{Z'}_{\rho}
+ {6\over13}\theta_M [2v_d\xi_{Vd} + 3\xi_{Ad}]  \eq

\bq  {\delta A_u\over A_u} = {12\over5}\delta^{Z'}_{\rho}
+ {4\over5}\theta_M [3v_u\xi_{Vu} - \xi_{Au}]  \eq

\bq  {\delta A_d\over A_d} = {15\over52}\delta^{Z'}_{\rho}
+ {5\over26}\theta_M [3v_d\xi_{Vd} - 2\xi_{Ad}]  \eq

\noindent
Assuming universality with respect to the three families of quarks we
also get:

\bq  {\delta \Gamma_h\over \Gamma_h} = {89\over59}\delta^{Z'}_{\rho}
+ {3\over59}\theta_M [4v_u\xi_{Vu} +12\xi_{Au}
+12v_d\xi_{Vd} +18\xi_{Ad}]  \eq

We can solve this set of equations and express the $Z'$ couplings in
terms of $\theta_M$, $\delta^{Z'}_{\rho}$ and the experimental values for
the shifts to the observables. The values that we shall give below
will always correspond to the upper bound, eq.(3), for $|\theta_M|$,
with the two possible signs $\eta_M=\pm 1$
and to experimental data taken at two standard deviations.\par
\underline{Lepton couplings} are obtained as:

\bq  \xi_{Vl} = {1\over2v_1\theta_M}[\delta A_l-3\delta^{Z'}_{\rho}] \eq

\bq  \xi_{Al} = {1\over2\theta_M}[{\delta \Gamma_l\over \Gamma_l}
-\delta^{Z'}_{\rho}] \eq

The experimental measurement 
$\Gamma_l =83.93\pm0.14 MeV$  
agrees with the SM prediction involving the $\epsilon_i$
parameters which depend on $m_t$ and $M_H$, \cite{epsilon}.
Taking
$m_t=180 \pm 12 GeV$ and $M_H=65-1000 GeV$ we get at most a total
relative shift ${\delta \Gamma_l\over \Gamma_l}=\pm 3\times 10^{-3}$. 
Combining with
$\delta^{Z'}_{\rho}$ given in eq.(2) and 
the upper bound for $|\theta_M|$
in eq.(3) we obtain:
\bq  \eta_M \xi_{Al} \simeq (-0.2\pm0.5)({M_{Z'}\over 1 TeV})  \eq

Concerning $A_l$, there is a disagreement between the LEP average
$A_l(LEP)=0.147\pm0.004$ and the SLC result
$A_{LR}(SLD)=0.1551\pm0.004$, whereas the SM prediction is
$A_e(SM)=0.144\pm0.003$. We then consider both cases. Combining
these results with $\delta^{Z'}_{\rho}$ in eq.(A.8), we obtain:
\bq \eta_M \xi_{Vl} \simeq (-2.25 \pm 6.25)({M_{Z'}\over 1 TeV}) \ \ \
(LEP) \eq

\bq \eta_M \xi_{Vl} \simeq (+1.75 \pm 6.25)({M_{Z'}\over 1 TeV}) \ \ \
(SLC) \eq

\underline{b-quark couplings} are obtained from :

\bq  \xi_{Vb} = {1\over 30v_b\theta_M}
[{325\over 13}{\delta A_b\over  A_b}
+{10\delta \Gamma_b\over \Gamma_b}-5\delta^{Z'}_{\rho}] \eq

\bq  \xi_{Ab} = {1\over 10v_b\theta_M}[{-8\delta A_b\over  A_b}
+{5\delta \Gamma_b\over \Gamma_b}-5\delta^{Z'}_{\rho}] \eq

We used for the $b\bar b$ anomaly the shift 
${\delta \Gamma_b\over \Gamma_b} =+0.03\pm 0.008$, but for $A_b$ we
have different results from LEP and from SLC to be compared with the SM
result $A_b(SM)=0.934$. From $A^b_{FB}$ at LEP,
$A_b=0.916\pm0.034$, we obtain ${\delta A_b\over A_b} =-0.02\pm 0.04$ 
and :

\bq \eta_M \xi_{Vb} \simeq (-3.45 \pm 20.72)({M_{Z'}\over 1 TeV}) \ \ \
(LEP) \eq

\bq \eta_M \xi_{Ab} \simeq (+4.58 \pm 9.84)({M_{Z'}\over 1 TeV}) \ \ \
(LEP) \eq

 Using the SLD result, 
$A_b=0.841\pm0.053$, we obtain ${\delta A_b\over A_b} =-0.1\pm 0.05$ and

\bq \eta_M \xi_{Vb} \simeq (-24.24 \pm 25.98)({M_{Z'}\over 1 TeV}) \ \ \
(SLC) \eq

\bq \eta_M \xi_{Ab} \simeq (+14.54 \pm 12.47)({M_{Z'}\over 1 TeV}) \ \ \
(SLC) \eq

For \underline{c-quark couplings} the solutions are :

\bq  \xi_{Vc} = {1\over 10v_c\theta_M}[{15\over4}{\delta A_c\over  A_c}
+{5\over3}{\delta \Gamma_c\over \Gamma_c}
-{35\over3}\delta^{Z'}_{\rho}] \eq

\bq  \xi_{Ac} = {1\over 10\theta_m}[-{5\over4}{\delta A_c\over  A_c}
+{5\delta \Gamma_c\over \Gamma_c}-5\delta^{Z'}_{\rho}] \eq

Experimental data are less precise than for b-quarks. We have
$ {\delta \Gamma_c\over \Gamma_c} =-0.1\pm 0.05 $ but for the
asymmetry there is again a
discrepancy between LEP and SLC. At LEP, from $A^c_{FB}$, 
$A_c=0.67\pm0.06$, whereas at SLC $A_c=0.606\pm0.09$, to be compared
with the SM prediction $A_c=0.67\pm0.002$. So with
${\delta A_c\over A_c} =0\pm 0.1$ at LEP one obtains :

\bq \eta_M \xi_{Vc} \simeq (-6.94 \pm 26.60)({M_{Z'}\over 1 TeV}) \ \ \
(LEP) \eq

\bq \eta_M \xi_{Ab} \simeq (-7.88 \pm 8.46)({M_{Z'}\over 1 TeV}) \ \ \
(LEP) \eq

whereas with ${\delta A_c\over A_c} =-0.1\pm 0.15 $ at SLC :

\bq \eta_M \xi_{Vc} \simeq (-20.38 \pm 40.62)({M_{Z'}\over 1 TeV}) \ \ \
(SLC) \eq

\bq \eta_M \xi_{Ab} \simeq (-6.01 \pm 9.70)({M_{Z'}\over 1 TeV}) \ \ \
(SLC) \eq

\noindent
Note that all above results correspond to the upper bound, eq.(3), for
$|\theta_M|$ and to experimental data taken with two standard deviations.

 \newpage

\renewcommand{\theequation}{B.\arabic{equation}}
\setcounter{equation}{0} 

{\large {\bf Appendix B : Dijet invariant mass distribution 
in hadronic collisions.}}\\

\vspace{0.3cm}

The observable that we consider is the dijet invariant mass ($M_{JJ}$) 
distribution:

\bq \frac{d\sigma}{dM_{JJ}} =  \frac{M^2_{JJ}}{2S}\int_{-\eta}^{\eta} 
d {\eta}_1 \int_{{\eta}_{min}}^{{\eta}_{max}} d{\eta}_2 \sum_{ij} 
\frac{1}{\cosh^2({\eta}^{\star})} f_i(x_1,M^2) f_j(x_2,M^2)
\frac{d\sigma_{ij}}{d{\hat t}} \eq
\noindent
where the $f_i(x,M^2)$ are the parton distribution evoluted at scale
$M^2$; $\eta$ has been defined in Sect.3, ${\eta}_1$ and 
${\eta}_2$  are the pseudorapidities of jets 1
and 2, $\eta_{min}=max[-\eta, ln{M_{JJ}\over\sqrt{s}}-\eta_1]$,
$\eta_{max}=min[+\eta, -ln{M_{JJ}\over\sqrt{s}}-\eta_1]$, 
whereas $\frac{d\sigma_{ij}}{d{\hat t}}$ is the partonic cross
section for the subprocess $ij \rightarrow 2 jets$. The momenta
fractions carried by initial partons read:
\bq x_1 = \frac{M_{JJ}}{\sqrt S} \exp({\eta}_B)  \eq
and 
\bq x_2 = \frac{M_{JJ}}{\sqrt S} \exp({-\eta}_B)  \eq
where ${\eta}_B = \frac{{\eta}_1 + {\eta}_2}{2}$.

The expression for the partonic cross sections can be found in
\cite{BGS}. The pure QCD terms for $gg \rightarrow gg$,  
$qg \rightarrow qg$,  $gg \rightarrow q \bar q$, $q \bar q \rightarrow
gg$ as well as the QCD and $\gamma$, $Z$ and $W$ exchange
contributions to the subprocess $qq \rightarrow qq$ are given in
eqs.(A1)-(A6) of \cite{BGS}. The subprocess 
$q \bar q \rightarrow q \bar q $ is obtained by performing the
crossing $s \leftrightarrow u$.
The QCD and W,Z,$\gamma$ exchange contributions to 
$qq' \rightarrow qq'$ are given by eqs.(A7)-(A14) of \cite{BGS}. By
crossing $s \leftrightarrow u$ one obtains the
$q \bar q' \rightarrow q \bar q'$ subprocess and by crossing 
$s \leftrightarrow t$ and then $t \leftrightarrow u$
 the $q \bar q \rightarrow q' \bar q'$ 
subprocess.
One has also to add the pure W exchange processes involving four
distinct quarks:
 $qq' \rightarrow q''q'''$, $q \bar q''' \rightarrow q'' \bar q'$, as
given by eqs.(A15) and (A16) of \cite{BGS}.  \par 
We have now to add the $Z'$ contribution to these various subprocesses.
The $Z'Z'$, $Z' \gamma$, $Z'W$ and $Z'g$ squared matrix elements can
be directly obtained from the  $ZZ$, $Z \gamma$, $ZW$ and $Zg$ ones
given in \cite{BGS}, by performing the replacement of $g_{Vq}$ by 
$\xi_{Vq} g_{Vq}$ and of $g_{Aq}$ by $\xi_{Aq} g_{Aq}$. More
precisely one has to replace the $C_L$ and $C_R$ $Z$ couplings to
left-handed and right-handed quarks by the following ones:

 \bq C'_{q,L} = \frac{1}{2} (g'_{Vq} + g'_{Aq}) = 
 \frac{1}{2} (\xi_{Vq} g_{Vq} + \xi_{Aq} g_{Aq}) \eq 

\bq C'_{q,R} = \frac{1}{2} (g'_{Vq} - g'_{Aq}) = 
 \frac{1}{2} (\xi_{Vq} g_{Vq} - \xi_{Aq} g_{Aq}) \eq 

The contribution due to the interference between the $Z$ and the $Z'$
is the only one that cannot be directly read off from their
expressions. We have computed it explicitely.
For the subprocess $qq \rightarrow qq$ we obtain (using the same
notations as in \cite{BGS}) :

 \bqa && T_{ZZ'} = 2 \alpha^2_Z \lbrack s^2 (\frac{1}{t_Z t_{Z'}} + 
\frac{1}{u_Z u_{Z'}} +\frac{1}{3} (\frac{1}{t_Z u_{Z'}} 
 + \frac{1}{u_Z t_{Z'}})) (C_{q,L}^2 C'^{\ 2}_{q,L} +
 C_{q,R}^2 C'^{\ 2}_{q,R})
\nonumber\\ 
 && + 2 C_{q,L} C'_{q,L} C_{q,R} C'_{q,R} (\frac{u^2}{t_Z
t_{Z'}}+  \frac{t^2}{u_Z u_{Z'}}) \rbrack \eqa

For the subprocess $qq' \rightarrow qq'$ we obtain:

 \bqa && T_{ZZ'} = 2 \alpha^2_Z \lbrack \frac{s^2}{t_Z t_{Z'}} 
(C_{q,L} C'_{q,L} C_{q',L} C'_{q',L}+C_{q,R} C'_{q,R} C_{q',R}
C'_{q',R}) \nonumber\\ 
&& + \frac{u^2}{t_Z t_{Z'}} (C_{q,L} C'_{q,L} C_{q',R} C'_{q',R} +
 C_{q,R} C'_{q,R} C_{q',L} C'_{q',L})  \rbrack \eqa

For subprocesses involving antiquarks the same crossings as previously
given have to be performed.

The complete expression for $\frac{d\sigma_{ij}}{d{\hat t}}$ is then
obtained by summing over the quark flavours (we have not considered
top production since its decay involves also a W leading to a
different topology) and adding to 
$\frac{d\sigma_{ij}}{d{\hat t}}(s,t,u)$ the crossed contribution  
$\frac{d\sigma_{ij}}{d{\hat t}}(s,u,t)$ due to the indiscernability of
jets.

\newpage

\centerline {\ {\bf Figure Captions }}

Fig.1 Fractional difference between dijet CDF data \cite{CDF} and QCD, 
compared to a hadrophilic $Z'$ of mass $M_{Z'}=800$ GeV for 
$\xi_{Vb}=4$, $\xi_{Ab}=3$, $\xi_{Vc}=4$ and $\xi_{Ac}=3$.\\

 Fig.2 Fractional difference between dijet CDF data \cite{CDF} and QCD, 
compared to a hadrophilic $Z'$ of mass $M_{Z'}=900$ GeV for 
$\xi_{Vb}=4$, $\xi_{Ab}=3$, $\xi_{Vc}=4$ and $\xi_{Ac}=3$.\\
 
Fig.3 Domains in $Z'bb$ vector and axial coupling ratios scaled by the
factor $(M_{Z'}/1TeV)$. Observability
limits from $\sigma_b$ at LEP2 with two possible accuracies,
5\% (Central dark),  10\% (Central grey). Upper and lower rectangles
correspond to the more restrictive SLC suggested domains, 
eq.~(\ref{lep}) and  eq.~(\ref{slc}).\\

Fig.4 Domains in $Z'bb$ vector and axial coupling ratios scaled by the
factor $(M_{Z'}/1TeV)$. Observability
limits from $A^b_{FB}$ at LEP2 with two possible accuracies,
5\% (Central dark),  10\% (Central grey). Upper and lower rectangles
correspond to the more restrictive SLC suggested domains, 
 eq.~(\ref{lep}) and  eq.~(\ref{slc}).\\

Fig.5 Domains in $Z'cc$ vector and axial coupling ratios scaled by the
factor $(M_{Z'}/1TeV)$. Constraint due to the $Z'bb$-$Z'cc$
correlation, eq.~(\ref{correl}) and the observability
of a 5\% effect on $\sigma_{had}$ at LEP2, for $|\xi_{Vb}|<2$, 
$|\xi_{Ab}|<1.5$ (white domain), for $|\xi_{Vb}|<4$, 
$|\xi_{Ab}|<3$ (grey + white domain).\\

\newpage
\thispagestyle{empty}
\[\null \hspace{-2cm}
\epsffile[22 24 664 746]{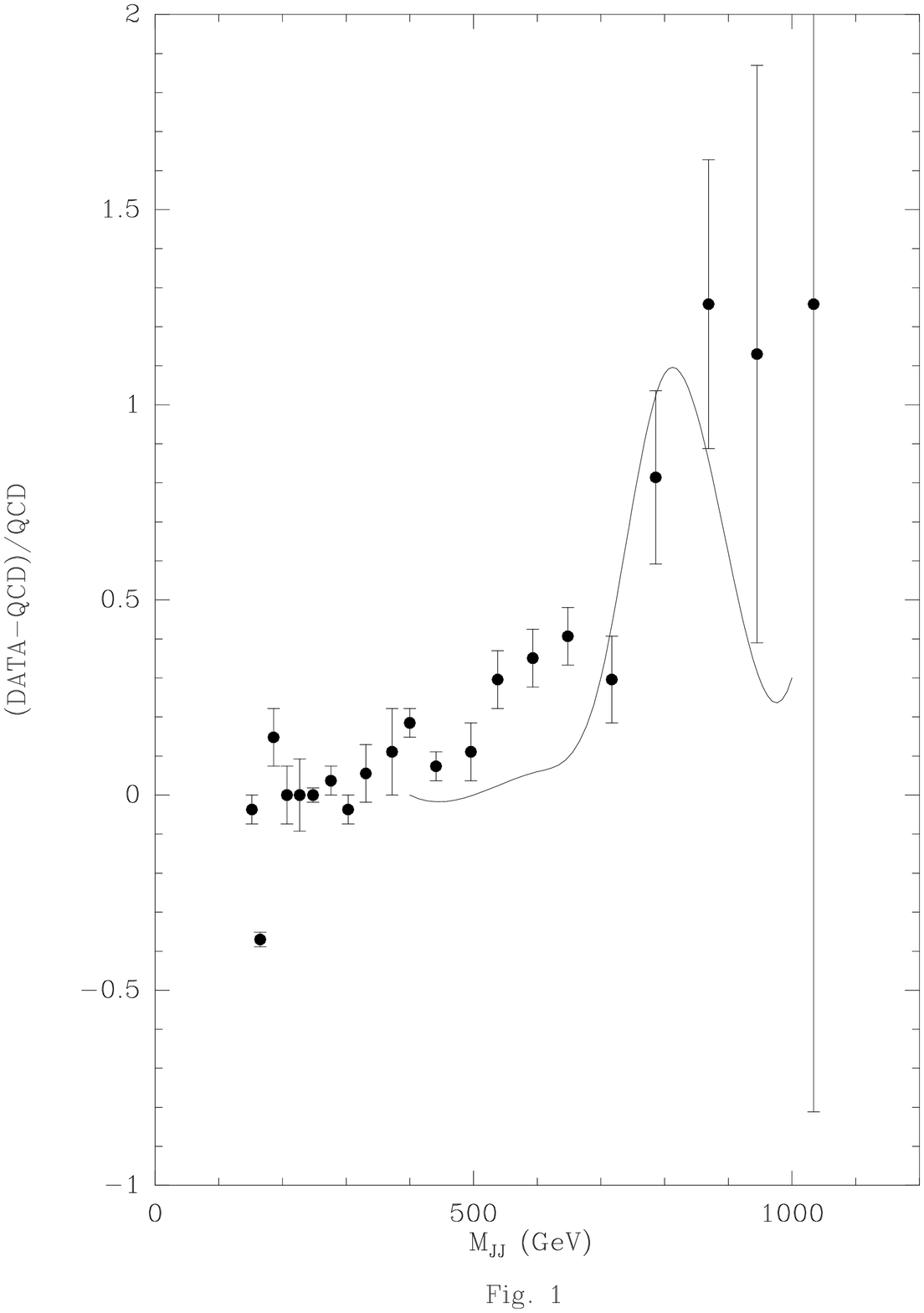}
\]
\newpage
\thispagestyle{empty}
\[ \null \null \hspace{-2cm}
\epsffile[22 24 664 746]{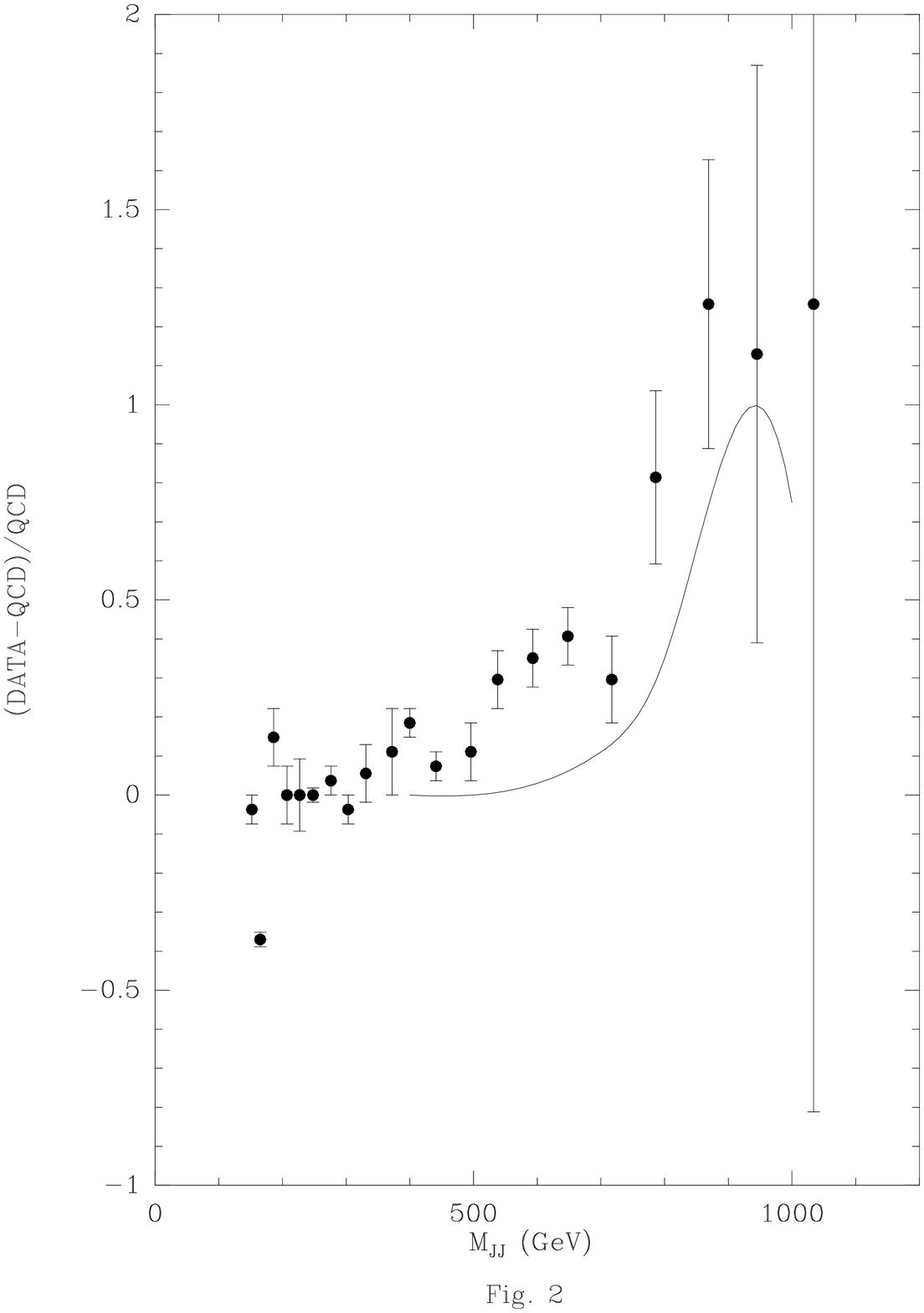}
\]
%
%
\newpage
\thispagestyle{empty}
\[
\epsffile[97 135 500 715]{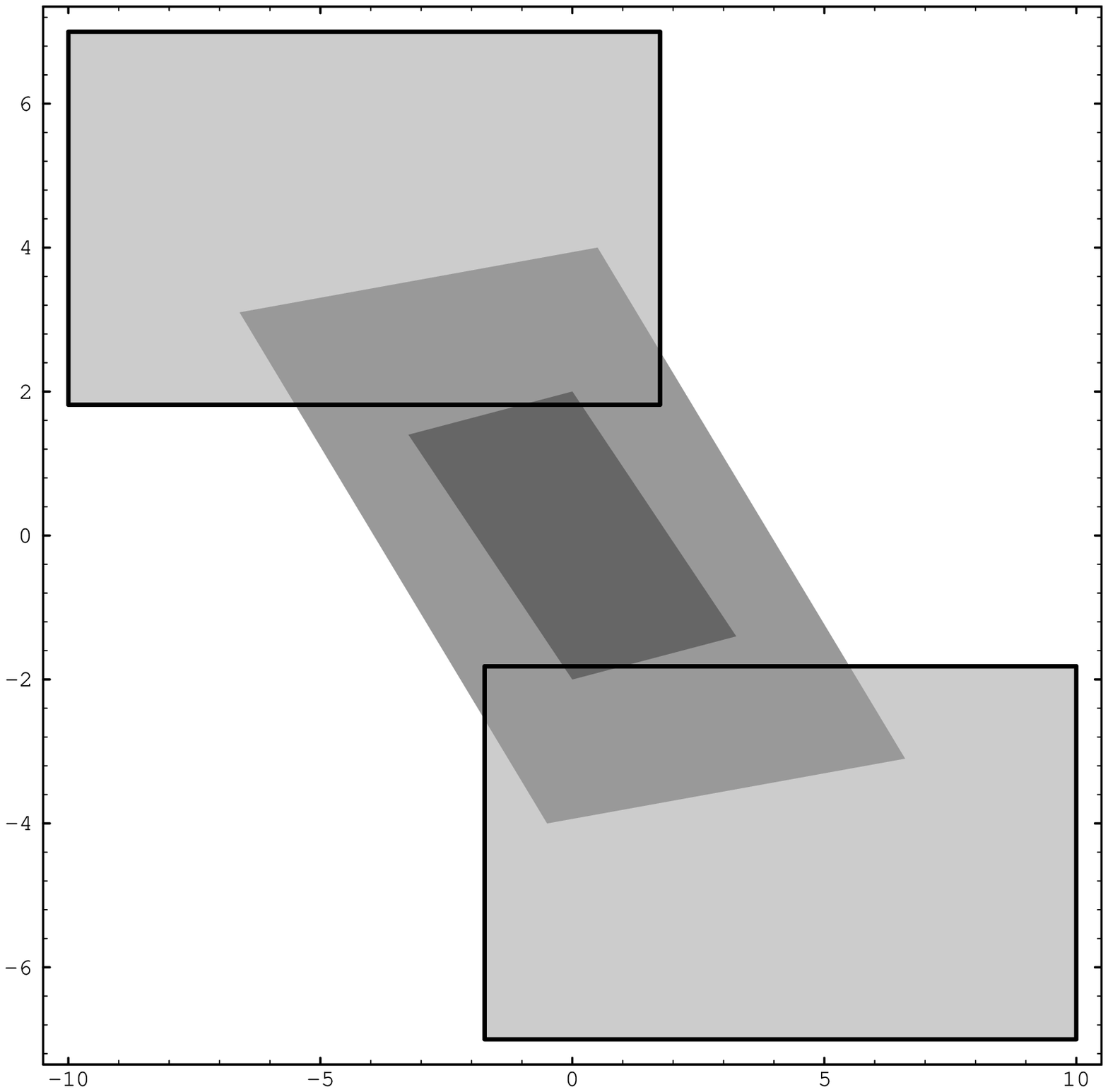}
\]
\centerline{ {\Large {\bf  Fig 3 }}}
\Large{
\vspace{-19.cm}\\
\null\hspace{-0.5cm}
 $\xi_{Ab}$
\vspace{15.2cm}\\
\null \hfill  $\xi_{Vb}$}
\newpage
\thispagestyle{empty}
\[
\epsffile[97 135 500 715]{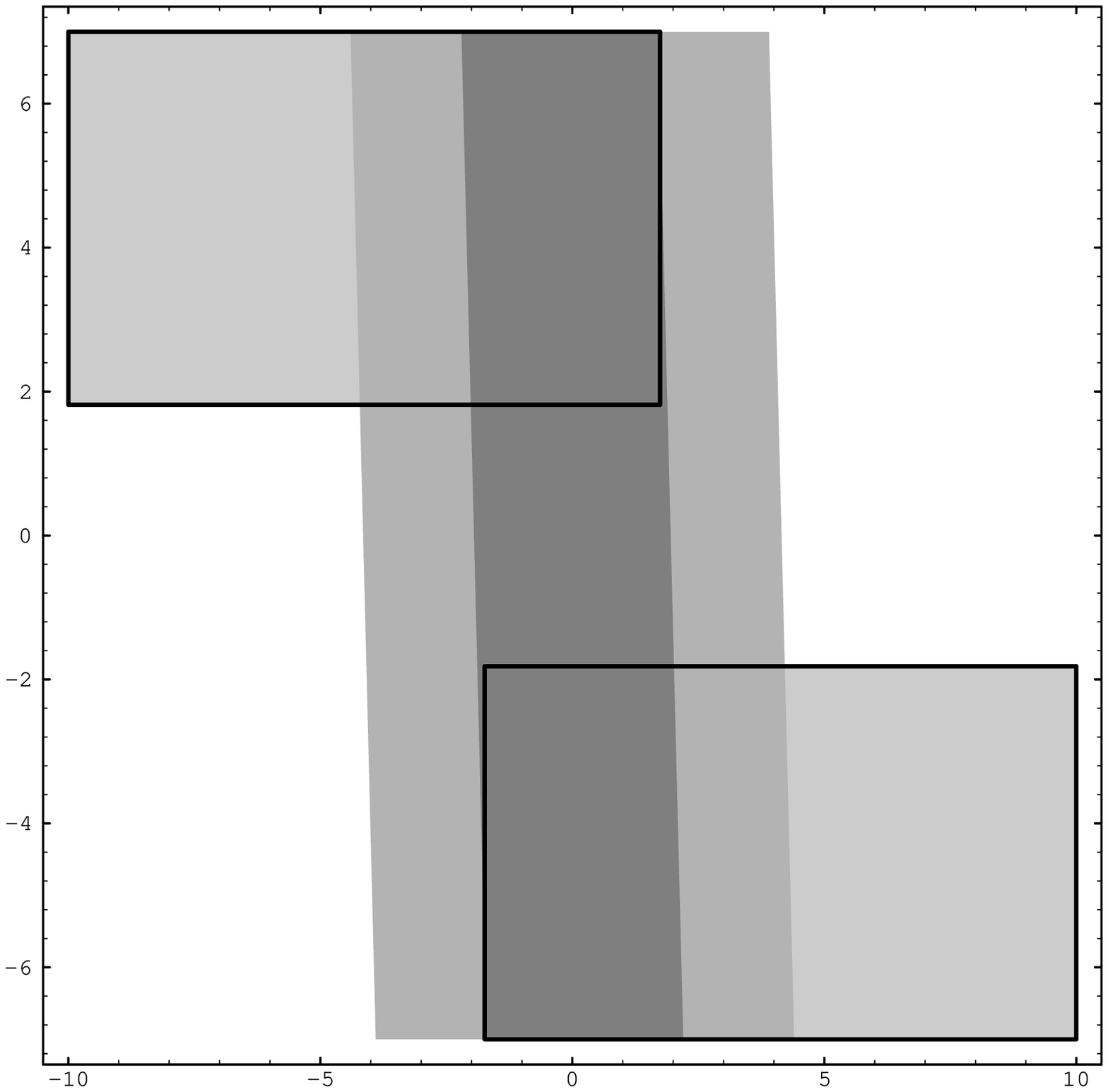}
\]
\centerline{ {\Large {\bf  Fig 4 }}}
\Large{
\vspace{-19.cm}\\
\null\hspace{-0.5cm}
 $\xi_{Ab}$
\vspace{15.2cm}\\
\null \hfill  $\xi_{Vb}$}
\newpage
\thispagestyle{empty}
\[
\epsffile[97 135 500 715]{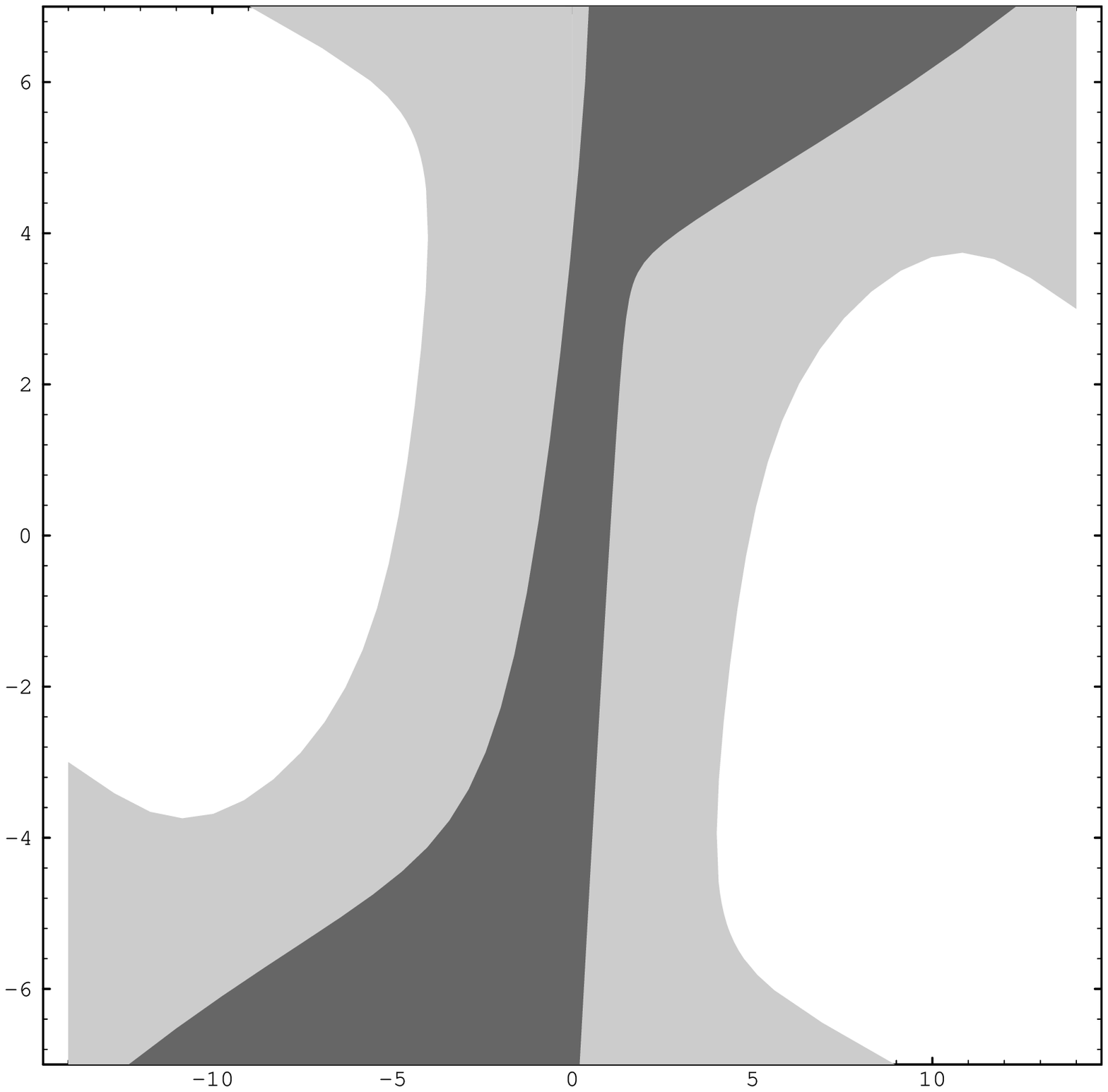}
\]
\centerline{ {\Large {\bf  Fig 5 }}}
\Large{
\vspace{-19.cm}\\
\null\hspace{-1.cm}
 $\xi_{Ac}$
\vspace{15.2cm}\\
\null \hfill  $\xi_{Vc}$}

\end{document}